# Is 40 the new 60? How popular media portrays the employability of older software developers

Sebastian Baltes,[1] George Park,[2] and Alexander Serebrenik[2]
[1]The University of Adelaide, Australia   [2]Eindhoven University of Technology, The Netherlands

*Abstract:*
Alerted by our previous research as well as media reports and discussions in online forums about ageism in the software industry, we set out to study the public discourse around age and software development. With a focus on the USA, we analyzed popular online articles and related discussions on Hacker News through the lens of (perceived) employability issues and potential mitigation strategies. Besides rather controversial strategies such as disguising age-related aspects in résumés or undergoing plastic surgeries to appear young, we highlight the importance of keeping up-to-date, specializing in certain tasks or technologies, and present role transitions as a way forward for veteran developers. With this article, we want to build awareness among decision makers in software projects to help them anticipate and mitigate challenges that their older employees may face.

## Introduction

Employability is a common concern among software developers. While recent graduates and career jumpers want to land their first job, more senior developers move between roles, teams, or companies. Recent media reports identified ageism as a major barrier in the hiring process of large software companies (https://bit.ly/2RKJW6S). This media coverage, statements such as Mark Zuckerberg's infamous opinion that "young people are just smarter" (https://bit.ly/2RIit5L), as well as discussions in popular online forums (https://bit.ly/2yfZGYN) all contribute to the notion that it is hard to be "old" and a successful software developer. This stereotype influences current and prospective developers alike—the power of language and its influence on individuals cannot be underestimated [1].

While the perception that older people are worse at learning new things is stable across cultures, age is usually also associated with wisdom and knowledge [2]. However, in software development, experience does not seem to be valued to the same degree as being fast and keeping up with the latest trends [3]. In this article, we followed up on that observation by studying which employability strategies popular online articles and related online discussions suggest for older software developers.

We start by qualitatively analyzing 24 popular online articles. We contextualize our findings using scientific literature on ageing in software development and a follow-up study investigating discussions about the above-mentioned articles on Hacker News.

We discuss our findings through the lens of (perceived) employability issues mentioned in the articles and forum discussions. We target decision makers in software projects such as managers and team leaders to alert them about potential issues their "veteran" employees may face, helping them to anticipate and potentially mitigate those issues. Besides implications for practitioners, we hope that this article triggers further research on the specific challenges of older software developers. Especially in industrialized countries, the demographic change leads to an older workforce. Nevertheless, the challenges that older developers face in a competitive field like software development are yet to be explored.

## Background

Societies across the world are ageing, a phenomenon that the United Nations consider "one of the most significant social transformations of the twenty-first century" (https://bit.ly/2K7tolx). Consequently, the number of older software developers is increasing. In the USA, for example, the number of software developers aged 55 to 64 years increased from 87,000 (8.3%) in 2011 to 195,000 (10.7%) in 2019 [4].

Davidson et al. [6] found that the top motivations for older developers were intrinsic, including community identification, internal values, and altruism. Younger developers mentioned career benefits and learning as major motivating factors. Morrison et al. suggest that older developers are less motivated by social interactions such as participating in chat rooms [5]. This shift in motivation is often caused by changing responsibilities, for example for family members, which again cause conflicts with work-related demands such as working overtime or keeping up-to-date [7, 9].

Moreover, in software development, being able to keep up-to-date and continuously learn is often linked to youth [7]. Consequently, older developers are considered less adaptive to change. However, older experts are also known to develop "domain-specific compensatory skills" to counter age-related cognitive decline, allowing them to maintain their performance [12].

In the career path of developers, moving to a non-programming role is seen as a fundamental step [11]. Thus, expressing a wish to continue as a developer is associated with a lack of drive. Xia and Kleiner point to another perception of older developers: they are supposedly more expensive due to higher pay and required medical support [8].

## Analyzing the Public Discourse

To analyze the public discourse on age and software development, we conducted a content analysis of popular online articles and discussions around those articles. First, we retrieved the top 100 results for the search query "age software developer" using the Google Custom Search API. We searched for "software developer" and not "software", because we were interested in the people developing software, not software systems in general. Further, we added "age" instead of "old", because we wanted to openly explore the discourse on age and software development without limiting ourselves to older developers from the beginning. We did, however, later check the first five relevant articles from the alternative result list and found them to be in line with our findings. We configured the search to only return English results, because most global online media targeting software development publish in English. To yield consistent search results regardless of our location, we configured the geolocation to be USA, because it is a large English-speaking country with a highly developed software industry having a significant proportion of older developers [4]. We executed the query with the above-mentioned configuration on August 2, 2019. The query results are available online (https://bit.ly/2z19eXW) together with our retrieval tool (https://bit.ly/2XEuWez).

Starting with the most popular results (according to Google's ranking), we followed all links and saved their content if the result was an online article either published on a news website (e.g., TechCrunch) or a blog (e.g., hosted on Medium). The articles and blog posts had to contain at least some editorial content and not just a collection of statistics or small contributions as in forum posts. We further excluded 24 (job) advertisements present in the result list. Of the inspected 100 links, 24 met our inclusion criteria (identified as A1-A24 in the published dataset).

For the included articles, we extracted metadata about the authors and publication media together with all textual content. In multiple—initial and focused—coding iterations, all three authors labeled the paragraphs of the articles independently while discussing the codes until agreeing on a common coding schema. The resulting schema captured different aspects of the public discourse on age and software development, including whose voices are reflected, whether age is described as being limiting, and which age-related aspects are reported. After finishing the coding, we observed that *employability* was a common theme throughout the articles. Therefore, two authors went over all articles again, focusing on suggested employability strategies for older developers—both from the perspective of an individual and from the perspective of companies. We collected and organized the identified strategies in a mind map (see Figure 1).

To not limit our view to the opinions captured in the online articles we analyzed, we extended our data collection to online discussions around those articles. We selected Hacker News as our target platform, because it is very popular within the software development community.

We searched for the titles of the included articles on the Hacker News website and identified 13 discussions around eight of the articles. We retrieved all 2099 comments from those discussions. Then, we normalized the content of the comments as well the content of all article paragraphs related to three frequently mentioned employability strategies: *mastering modern technologies* (mentioned in ten articles), *moving to a management role* (six articles), and *specialization* (five articles). Normalization included converting the content to lower case, removing special characters, stop words, as well as stemming the words. We tokenized the normalized comments and paragraphs into 3-grams and calculated their pairwise cosine similarity considering the BM15 weighting scheme. We then selected the five most similar comments for each paragraph, resulting in 105 paragraph-comment pairs. Two of the authors coded different subsets with an overlap of 17 pairs. The coders determined whether the comments agree or disagree with the article paragraphs.

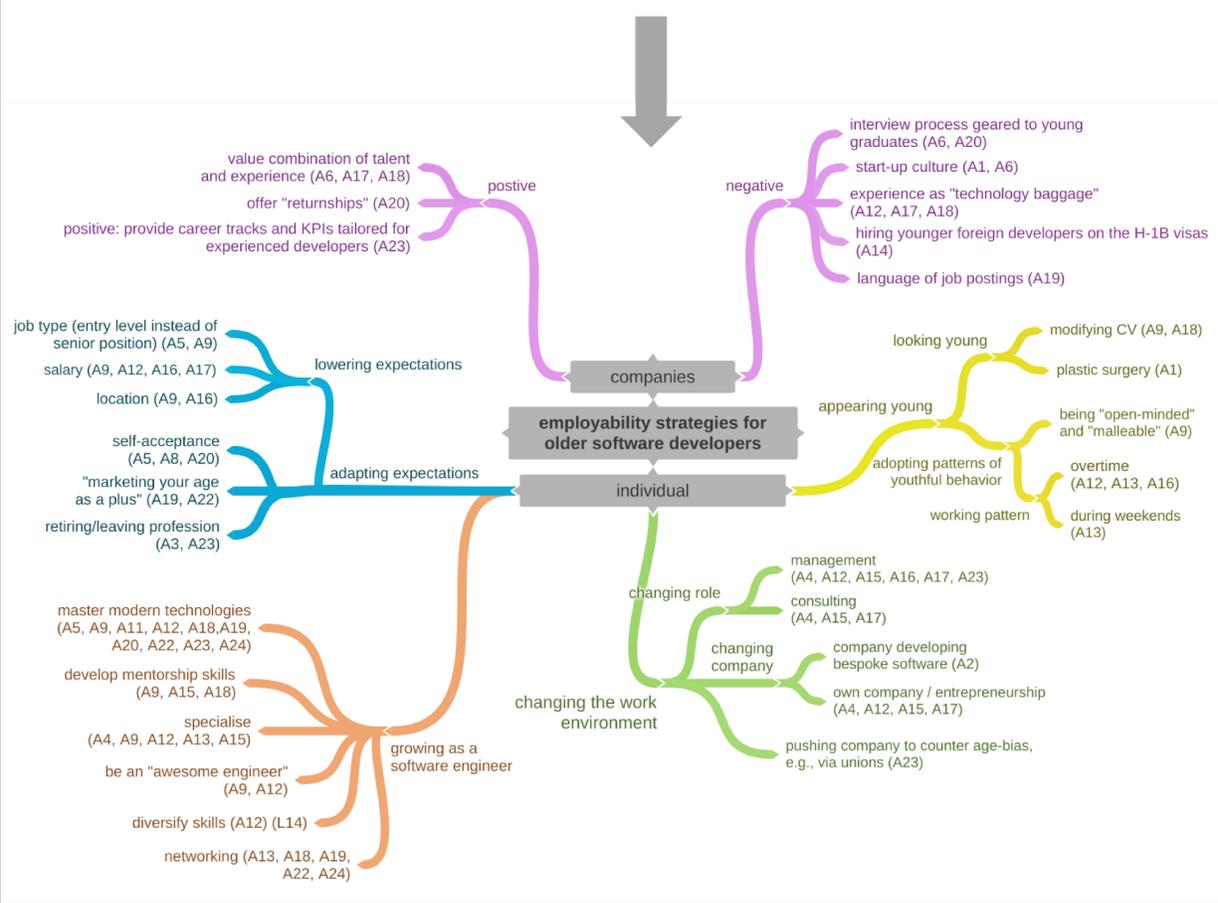

Figure 1: The twenty-four articles considered in this study and the mind map based on those articles structuring mentioned employability strategies for older software developers.

Table 1: Age definitions in the analyzed articles.

| An "old" software developer is... | Number of articles |
|---|---:|
| 30+ | 4 |
| 35+ | 3 |
| 35-40+ | 2 |
| 40+ | 7 |
| 45+ | 2 |
| 50+ | 1 |
| Not stated explicitly | 5 |

## Reported Age Definitions and Employability Strategies

We started by collecting the authors' **definition for being an "old" software developer** (see Table 1). Surprisingly, the most common definition was 40+ years, followed by 30+ years. One reason for the 40+ definition could be that 40 is the threshold for protection under the US Age Discrimination in Employment Act (https://bit.ly/3agASwY). Still, even the highest number, 50+ years, which is a common threshold in scientific papers [6], is far away from a typical retirement age in industrialized countries. Even though our sample size is relatively small, this strongly suggests a very biased notion of "age" or "being old" in the public discourse around age and software development. Another interesting observation is captured in article A8, where the author lists 47 questions on Quora about being too old to become a software developer, ranging from age 14 to age 60. This further suggests that the fear of being too old to work in the software industry is common among a wide range of age groups.

The most commonly suggested **employability strategies** for older developers were *specialization*, for example in software architecture or legacy code, *networking*, *moving to a management role*, and *mastering modern technologies*. The latter two aspects have already been discussed in related work [7]. However, in particular the importance of mastering modern technologies is quite controversial among developers on Hacker News: while some commenters indicate the importance of keeping up to date with technology development, others stress that learning new technology cannot counter the "cultural mismatch" but also that there "are skills past simply becoming proficient in new tools". Developers were more in agreement with transition to management as a viable strategy: they see management as both "safe" and breaking the salary ceiling. There were, however, also references to dedicated tracks for developers who do not want to make that transition (mentioned in A23). We grouped those strategies, together with related aspects such as *diversifying skills* and *networking*, into the high-level categories *growing as a software engineer* and *changing the work environment*.

While those strategies come with their own challenges, we want to focus on novel aspects not yet covered in related work. First of all, the high-level category *adapting expectation* covers strategies based on the premise of "you're old, get over it" (A5). The category's spectrum ranges from *marketing your age as a plus* over *self-acceptance* and *lowering expectations* in terms of salary, seniority level, and location, to *retirement*.

The second category we want to focus on in this article is *appearing young*, which captures controversial strategies such as *modifying your résumé* to disguise age-related aspects as well as u*ndergoing plastic surgery* to look younger, an aspect that has recently been picked up by major US news outlets (https://wapo.st/34J9d6F). This category further contains *adopting patterns of youthful behavior*, including *working overtime* or *during weekends*, which are strategies known to conflict with other responsibilities such as family [7, 9].

The above-mentioned strategies are centered around the individual developer. We did, however, also observe strategies related to *company culture and practices*. Some articles mentioned that companies should explicitly value both talent and experience. An extreme example for not following this advice is to negatively frame experience as "technology baggage" (A12, A17, A18). A positive strategy is presented in article A20, where the author reports on so-called "returnships" that help older developers to re-join software development after a career break. Unfortunately, most other strategies were negative examples of what companies may do to discourage older developers from joining or staying in the company. Beside not valuing experience, the language used in job postings as well as the often-related "startup culture" with its cultural norms and social events, together with an interview process focusing on content that recent graduates learned [10], all discourage older developers.

## Conclusion

"Am I too old to be a developer" reverberates through different media involving software developers, from Quora posts to Hacker News, and from online magazines to scientific papers. In this study, we have focused on the employability strategies recommended to "older" software developers, based on a qualitative analysis of US-focused articles. While not all observations may be transferable to other countries, the underlying problem is not limited to the USA (https://bit.ly/2Bcuc7y). We observed that already the interpretation of "old" is strongly biased in the public discourse around age and software development.

Online articles provide a plethora of recommendations on employability strategies. While moving to a management role, mastering modern technologies, and specialization are most frequently mentioned, we also observe rather controversial strategies such as disguising age-related aspects in résumés or undergoing plastic surgeries to appear young, as well as strategies requiring developers to accept their age. Against this background, we argue that preserving employability cannot and should not solely be the burden of the developers themselves. Companies should take their share of responsibility, encourage practices that welcome developers of any age and curb those negatively affecting older developers. Examples for such negative practices include glorifying working late or on weekends, advertising a "startup culture" that may discourage older developers from applying, and tailoring the interview process to recent graduates.